%% file: khs1.tex
\input{epsf}
\documentstyle[prd,aps,twocolumn]{revtex}
\input{title2.tex}

\begin{document}
\draft
\preprint{LA-UR 97-2234, IOP-BBSR/97-XXXX}

\title{Exact Thermodynamics of the Double sinh-Gordon Theory in 
1+1-Dimensions}
\author{Avinash Khare,$^{1,*}$ Salman Habib,$^{2,\dagger}$ and Avadh 
Saxena$^{3,\#}$}
\address{$^1$Institute of Physics, Sachivalaya Marg, Bhubaneswar 751
005, India} 
\address{$^2$T-8, MS B285, Theoretical Division, Los Alamos National
Laboratory, Los Alamos, New Mexico 87545}
\address{$^3$T-11, MS B262, Theoretical Division, Los Alamos National
Laboratory, Los Alamos, New Mexico 87545}

\abstract
{We study the classical thermodynamics of a $1+1$-dimensional
double-well sinh-Gordon theory. Remarkably, the Schr\"odinger-like
equation resulting from the transfer integral method is quasi-exactly
solvable at several temperatures. This allows exact calculation of the 
partition function and some correlation functions above and below the 
short-range order (``kink'') transition, in striking agreement with
high resolution Langevin simulations. Interesting connections with 
the Landau-Ginzburg and double sine-Gordon models are also 
established.}
 

\maketitle2
\narrowtext

The statistical mechanics of nonlinear coherent structures in low
dimensions has long attracted theoretical attention, both for the
intrinsic interest in such fundamental problems as kink nucleation and
dynamics, as well as in diverse applications, {\em e.g.}, in
conducting polymer physics \cite{poly} and DNA denaturation
\cite{mp}. Both analytic and numerical techniques have been applied to
these problems: well known among them are the (analytic) transfer
integral method and the (numerical) Langevin method. The transfer
integral technique converts the problem of finding the classical
partition function $Z_{cl}$ to an eigenvalue problem for a
Schr\"odinger-like equation to which familiar approximation methods
such as WKB can then be applied \cite{ssf}. The advantage of Langevin
methods is that (unlike Monte Carlo) real time quantities such as
temporal correlation functions can be computed, and kinks/antikinks
tracked both in space and time.

In the past, applications of these methods
have yielded comparisons of approximate analytic results with
numerical data to only rather low levels of accuracy, of order of tens
of per cent \cite{fash}. In this Letter we report substantial progress
on both fronts. We discuss a nonintegrable $1+1$-dimensional field
theory for which thermodynamic quantities can be computed exactly at
several temperatures using techniques from quasi-exactly solvable
(QES) potentials \cite{razavy} in quantum mechanics. This theory is in
the same class as the more familiar Landau-Ginzburg model and also
admits exactly known kink solutions \cite{bk}. We have carried out
very high resolution Langevin simulations and find excellent agreement
with the exact results at the checkpoint temperatures. The high
accuracy of the Langevin simulations allows the use of the probability
distribution function (PDF) to directly compute thermodynamic
quantities \cite{pdf} thus providing an alternative to conventional
methods based on fluctuations.

In order to calculate $Z_{cl}$ exactly at various temperatures one has
to solve a Schr\"odinger equation with a temperature dependent
mass. While completely solvable potentials are rare, in
the last few years several double-well QES models have been discovered
for which the exact classical partition function can be found at one
given temperature. The drawback has been that the
exact eigenstates are only known for a given set of couplings and as a
result, it has not been possible to obtain the exact $Z_{cl}$ at more
temperatures. In this Letter we show that for the double
sinh-Gordon (DSHG) QES problem, if the ground state energy is known
for $n$ different values of coupling constants, then $Z_{cl}$ can be
evaluated for any of these theories (with a given set of coupling
constants) at $n$ different temperatures. This is also true for the
triple well $\phi^6$ and the double sine-Gordon (DSG) models,
results for which will be reported elsewhere \cite{hks}. We conjecture
that this result holds for a large class of QES problems.

The double-well $\phi^4$ model in $1+1$-dimensions has been
extensively studied. However, in this case the Schr\"odinger equation 
does not possess known exact solutions. To overcome this problem we turn 
to the DSHG potential: 
\begin{equation}
V_{DSHG}(\phi) = (\zeta\cosh 2\phi-n)^2~,
\label{vdshg}
\end{equation}
where $\zeta$ is a positive parameter. In order to have a double-well
potential, $n>\zeta$, in which case the two minima are located at 
$\cosh2\phi_0 = n/\zeta$. Moreover, for the system to be QES, $n$ has
to be a positive integer. This potential is the 
hyperbolic analog of the double sine-Gordon system.  Similar potentials 
arise in the context of the quantum theory of molecules ({\em e.g.} a 
homonuclear diatomic molecule), wave motion describing the normal modes 
of vibration of a stretched membrane of variable density, and as the 
solution of a Fokker-Planck equation \cite{refs}. The 
hyperbolic analog of the sine-Gordon equation is a single well potential 
(sinh-Gordon) and thus uninteresting from the soliton statistical 
mechanics perspective.

The DSHG potential written in the form (\ref{vdshg}) has all the
generic features of a double-well potential such as Landau-Ginzburg,
but allows for much greater analytic progress. Below we find exact
solutions for $1)$ a kink, $2)$ phonon dispersion, $3)$ a kink
lattice, and $4)$ the first few eigenvalues and eigenfunctions of the
transfer operator at certain temperatures, allowing thereby analytic 
calculation of the PDF and correlation functions in the thermodynamic 
limit. 

We exhibit below the exact kink and kink lattice solutions for the
DSHG theory (details will be given in Ref. \cite{hks}). A kink
is a time independent solution resulting from the minimization of the
total energy density $\varepsilon(x) = V_{DSHG}(\phi) + (g/2)\phi_x^2$
with the boundary conditions $\phi\rightarrow\pm\phi_0$ as
$x\rightarrow\pm\infty$. The constant $g$ is often introduced in
condensed matter treatments as a phenomenological parameter and
controls the kink size. In a field theoretic context, however, $g=1$,
and this is the value we choose here. (All the solutions given below
can be written for arbitrary $g$.)

The kink/antikink solution, located at $x_0$ is, $(n>\zeta)$
\begin{equation}
\phi(x) = \pm \tanh^{-1}\left(\tanh \phi_0
\tanh\left(\frac{x-x_0} {\xi}\right)\right) 
\end{equation}
where $\tanh \phi_0=\sqrt{(n-\zeta)/(n+\zeta)}$,
$\xi = [2(n^2-\zeta^2)]^{-1/2}$. The kink has topological charge
$Q = \int_{-\infty}^{\infty}\frac{\partial\phi}{\partial x}dx =
2\phi_0$. Traveling kink solutions are obtained by boosting to
velocity $v$ via $x \rightarrow (1-v^2)^{-1/2}(x-vt)$. The statistical
mechanics of kinks is governed largely by the kink energy (or rest
mass): 
$$
E_s = 4\xi n\sqrt{n^2-\zeta^2}\tanh^{-1}\left(\sqrt{\frac{n-\zeta}
{n+\zeta}}\right) - 2\xi (n^2-\zeta^2).  
$$

The phonon dispersion around the minima $\pm\phi_0$ for this model is
$\omega_q^2=q^2+8(n^2-\zeta^2)=q^2+(2/\xi)^2$ and the phonon contribution 
to the free energy per unit length is
\begin{equation}
F_{vib}={1\over 2\pi\delta}\ln\left({2\pi\over\delta\beta}\right)
+{1\over\beta}\sqrt{2(n^2-\zeta^2)}~,
\end{equation}
with $\delta$ being the lattice constant and $\beta\equiv 1/k_BT$.

In order to understand kink-antikink interactions, it is very useful to 
construct kink lattice solutions (a kink/antikink chain). For the DSHG 
theory, this solution is
\begin{equation}
\phi_L(x) =
\pm\tanh^{-1}\left(\tanh\phi_1~\hbox{sn}\left({x-x_0\over\xi_L},
k\right)\right)~,  \label{latt}
\end{equation}
$$
k = \frac{\tanh\phi_1}{\tanh\phi_2};~\xi_L = {k\over 2 \sqrt{2}\zeta
\sinh\phi_1\cosh\phi_2};~d = 4K\xi_L,  
$$
where $d$ is the periodicity of the kink lattice, $K(k)$ is the
complete elliptic integral of the first kind with modulus $k$,
sn$(x,k)$ is the Jacobi elliptic function, and with $0 < V_0 <
V(\phi=0)=(n-\zeta)^2$,
\begin{equation} 
\cosh2\phi_{1,2} = \frac{n}{\zeta}\mp\frac{\sqrt{V_0}}{\zeta};~k^2 =
\frac{n^2-(\sqrt{V_0}+\zeta)^2}{n^2-(\sqrt{V_0}-\zeta)^2}. 
\label{cosh} 
\end{equation}
The topological charge (per period) in the lattice problem $Q_L =
2\phi_L(K) = 2\phi_1$ is smaller than the single kink case. The kink
size in the kink lattice, $\xi_L$, is also smaller than the free kink
size $\xi$.

The energy of the kink lattice per period ({\em i.e.} energy per
kink-antikink pair plus the interaction energy) is
$$
E_L =4\xi_L\left[(n+\zeta)^2K+\frac{\xi^2}{\xi_L^2}(n^2-\zeta^2)(K-E)
-4n\zeta\Pi\right], 
$$
where $E(k)$ and $\Pi(\tanh^2\phi_1,k)$ are complete elliptic
integrals of second and third kind, respectively. In the
dilute limit ($k \rightarrow 1$, $d \rightarrow \infty$) the
divergences in $K(k)$ and $\Pi(\tanh^2\phi_1,k)$ exactly cancel out
and we recover the single kink result $E_s$. The interaction energy as
a function of separation ({\em i.e.} $k$ or $d$) is given by $E_{in} =
E_L-2E_s$.

Turning now to the computation of $Z_{cl}$, we note that this
calculation can be divided into two parts: a trivial Gaussian
integration over the field momentum, and a computation of the
configurational partition function, which via the transfer integral
method becomes equivalent to solving a Schr\"odinger-like equation
\cite{ssf}. The Hamiltonian for the DSHG theory is
\begin{equation}
H=\int dx\left[{1\over 2}\pi^2+{1\over
2}\left(\partial_x\phi\right)^2+V_{DSHG}(\phi)\right] 
\end{equation}
and this leads to the Schr\"odinger equation for the eigenvalues and
eigenfunctions of the transfer operator,
\begin{equation}
-{1\over 2\beta^2}{\partial^2\over\partial\phi^2}\Psi_k
+(\zeta\cosh 2\phi-n)^2\Psi_k=E_k\Psi_k
\label{seqn}
\end{equation}
Remarkably, this equation is an example of a QES system. Using results 
for a related potential from Ref. \cite{razavy}, at $2\beta^2 = 1$ the
eigenstates of the first $n$ levels can be found for $n =
1,2,3,4$. (We have extended this to the cases $n = 5,6$.)
However, what one really wants is to consider a given fixed-$n$ theory
and obtain eigenstates at {\em different} temperatures. It is easy to
see from Eq. (\ref{seqn}), by simple rescaling, that solutions of a
fixed-$n$ theory at certain values of $\beta$ are the same as the
solutions of {\em another} theory (different $n$ 
and $\zeta$) at $2\beta^2 = 1$. Depending on the chosen value of $n$,
exact solutions are available at different fixed values of
$\beta$. Here, we restrict ourselves to one such family $(n=2)$ which
allows the exact computation of the first few eigenstates at $8\beta^2
= m^2~(m=1,\cdots,6)$. For illustration, two examples of the
(unnormalized) ground states are given below (see also Fig. 1). The first 
(high temperature, $\beta^2=1/8$) has an eigenfunction with a single peak
while the second (lower temperature, $\beta^2=1/2$) has a double peak:
\begin{eqnarray}
\Psi_0(\phi)|_{\beta^2={1\over 8}}&=& \exp\left(-{1\over 4}\zeta 
\cosh 2\phi\right)~,\nonumber\\
\Psi_0(\phi)|_{\beta^2={1\over 2}}&=& \cosh\phi\exp\left(-{1\over 2}
\zeta \cosh 2\phi\right)~,
\label{exactgs}
\end{eqnarray}
with corresponding ground state energies, $E_0=1+\zeta^2/4$,
$E_0=\zeta^2-2\zeta+3$. The PDF for the field is the square of
the normalized ground state eigenfunctions. Solutions at higher
energies and other values of $\beta$ are given in Ref. \cite{hks}. 

Once the eigenvalues of the transfer operator are known, they can be
used to compute the correlation functions
$C_1(x)=\langle\phi(0)\phi(x)\rangle$ and
$C_2=\langle\delta\phi^2(0)\delta\phi^2(x)\rangle$, using
\begin{eqnarray}
C_1(x)&=&\sum_k |\langle\Psi_k|\phi|\Psi_0\rangle|^2
\exp\left[-\beta |x|(E_k-E_0)\right]~,
\label{c1}\\ 
C_2(x)&=&\sum_k |\langle\Psi_k|\delta\phi^2|\Psi_0\rangle |^2 
\exp\left[-\beta |x|(E_k-E_0)\right].  \label{c2}
\end{eqnarray}
It is apparent that at large distances, $C_1$ and $C_2$ are dominated
by the lowest state with nonvanishing matrix elements: the first
excited state in the case of $C_1$ and the second excited state in the
case of $C_2$. Since $E_0$, $E_1$, and $E_2$ are known at certain
temperatures, the large distance behavior of these correlation
functions can be found exactly and compared with the results from
simulations. Static structure factors may also be calculated in much
the same way.

At this point, it is important to mention the connection between the
``quantum'' calculations and kink physics. In the context of kink
statistical mechanics, it is usual to introduce a phenomenological
description of kinks as particles in a grand canonical
ensemble. However, this is unnecessary, and all such thermodynamical
information can be extracted directly from the Schr\"odinger
description of the transfer operator. For example, the kink density
has been obtained in this way in Ref. \cite{ahk}. Simpler quantities
like $C_1$ and $C_2$ have obvious natural interpretations in terms of
kinks. The $C_1$ correlation length is related to the kink/antikink
spacing and increases monotonically as $\beta$ increases (Fig. 2). 

The behavior of the $C_2$ correlation length requires a little 
explanation, since $C_2$ is not directly sensitive to domain size. At 
both high temperatures (no kinks) and low temperatures (number of kinks 
exponentially suppressed), the correlation length is essentially that 
set by thermal phonons, and is therefore small. However, at temperatures 
close to the kink transition, nonlinear fluctuations on the kink length 
scale become important and can dominate $C_2$. At these temperatures one 
expects the $C_2$ correlation length to rise to a maximum value, of
order the kink size, and this is indeed what we observe numerically 
(Fig. 3). The Schottky anomaly in the specific heat \cite{ahk} arises for 
the very same reason.

The exact results described above can be compared against those
obtained from Langevin methods. The additive noise Langevin equation
for the DSHG theory is $$
\partial_{tt}^2\phi=\partial_{xx}^2\phi-\eta\partial_t\phi-
4\zeta(\zeta\cosh 2\phi-2)\sinh 2\phi+F(x,t),
$$
where $F(x,t)$ is a stochastic (Gaussian, white) external force which
satisfies the fluctuation-dissipation relation linking the noise
strength to the viscosity $\eta$,
\begin{equation}
\langle F (x,t) F (x',t') \rangle = 2\eta\beta^{-1}\delta (x-x')
\delta (t-t').
\end{equation}
This stochastic PDE can be solved by standard methods \cite{stand}
which we implemented on massively parallel computers. Typical choices
for the lattice constant are often dictated by memory limitations
rather than by accuracy. Comparisons with the exact results have led
us to conclude that errors up to $30\%$ may be expected if lattice
discretization is done as coarsely as has been the norm so far in
numerical calculations. The existence of nontrivial exact continuum
results has proven to be essential in carefully estimating error and
convergence in field theoretic Langevin simulations \cite{hl}. Our
present simulations were typically performed on $5\times10^5$ site
lattices with a lattice constant $\delta=0.025$ and time-step
$\epsilon=.005$.

Fig. 1 shows the striking agreement between the numerically obtained
and the exact continuum PDFs at three temperatures: The worst case
departure is at the level of parts per thousand. The comparisons for
the inverse correlation lengths are given in Figs. 2 and 3. DSHG
system parameters are $n=2$, $\zeta=0.05$. For $C_1$, the numerical
values are $1/\lambda=0.1425$ ($\beta^2=1/2$) and $1/\lambda=0.012$
($\beta^2=9/8$) as compared to the exact values in the continuum
theory of $0.14142$ and $.0105$, respectively. The small offset
between the continuum and lattice calculations is due to the finite
value of the lattice constant and is consistent with estimates from
higher-order contributions to the transfer integral \cite{hl}.

\vspace{.5cm}
\epsfxsize=8.0cm
\epsfysize=4.5cm
\centerline{\epsfbox{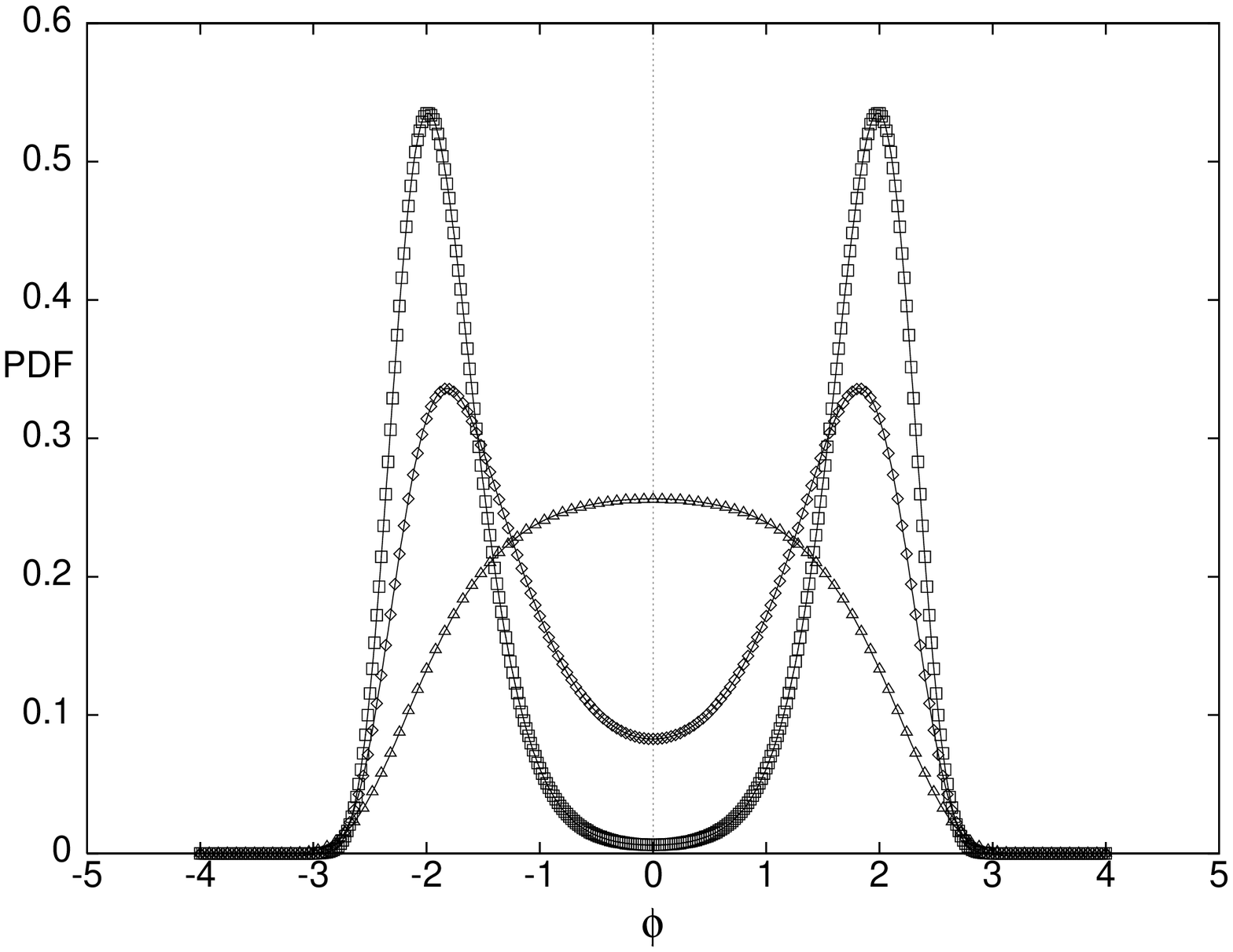}}
\vspace{.25cm}
{FIG. 1. {\small{The numerically evaluated PDFs at three values of
$\beta^2$: 1/8 (triangles), 1/2 (diamonds), and 9/8 (squares). The
corresponding continuum exact solutions are the solid lines.}}}\\ 

\vspace{.2cm}
\epsfxsize=7.5cm
\epsfysize=4.5cm
\centerline{\epsfbox{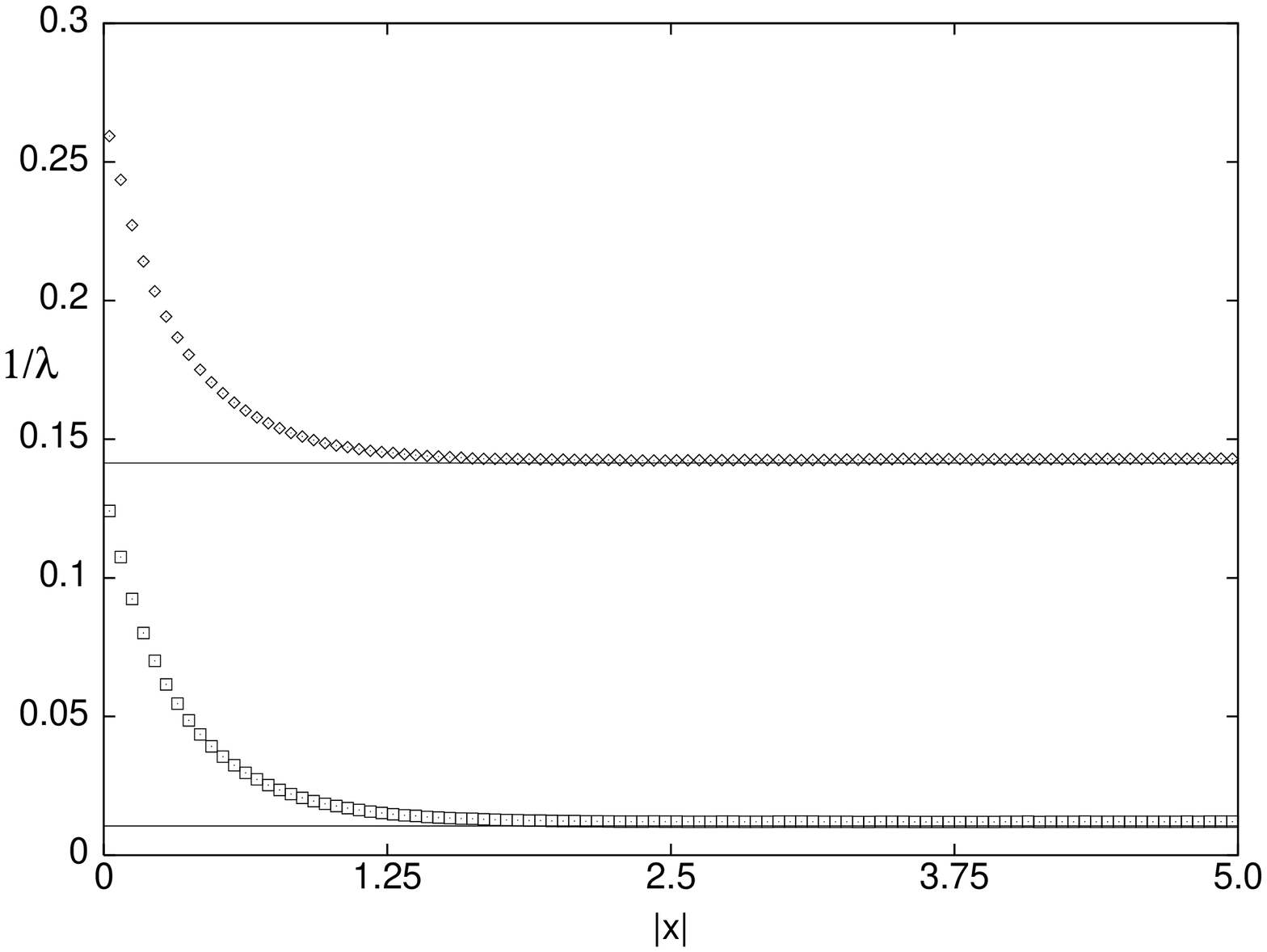}}
\vspace{.25cm}
{FIG. 2. {\small{The numerically obtained inverse correlation lengths
from $C_1(x)$ for $\beta^2$: 1/2 (diamonds), and 9/8 (squares). The
large $|x|$ continuum exact results are the solid lines.}}}\\

\vspace{.2cm}
\epsfxsize=7.5cm
\epsfysize=4.5cm
\centerline{\epsfbox{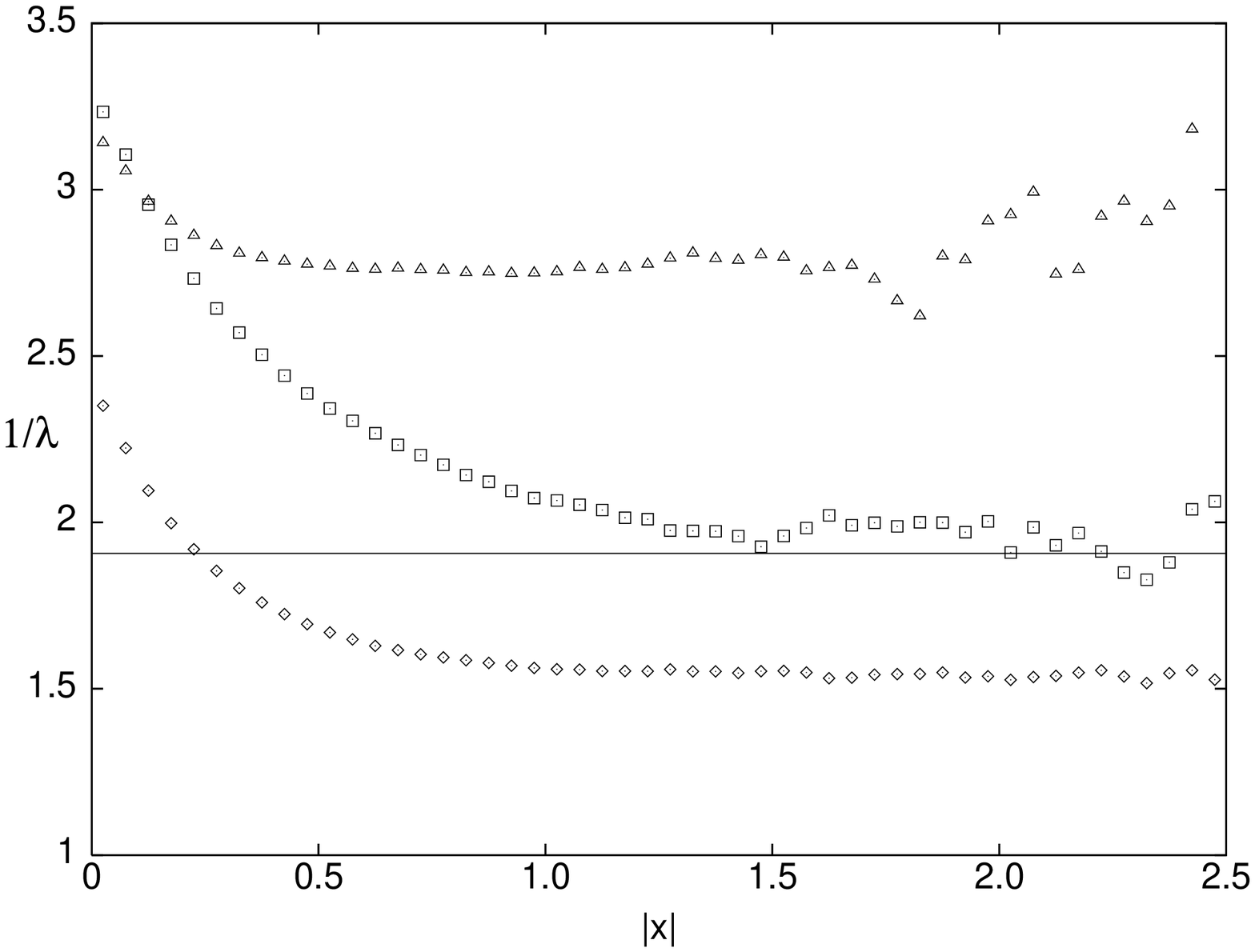}}
\vspace{.25cm}
{FIG. 3. {\small{The numerically obtained inverse correlation lengths
from $C_2(x)$ for the same three temperatures and with the same
conventions as in Fig. 1. The solid line is the large $|x|$ continuum
exact result for $\beta^2=9/8$. The largest correlation length is at
the intermediate value of $\beta$ (see text).}}}\\

The high quality of these numerical simulations implies that the PDF 
can now be used directly to compute thermodynamic quantities at {\em any}
temperature. Since the PDF is just the square of the ground state wave
function of the Schr\"odinger equation (\ref{seqn}), one can use it to
compute the ground state energy $E_0$ numerically, from which the
internal energy ($U=\partial E_0/\partial \beta$), the free energy
($F=E_0/\beta$), and the entropy ($S=\beta \partial
E_0/\partial\beta-E_0$) can all be computed in a straightforward
way \cite{hks}. The specific heat involves two $\beta$ derivatives and is
difficult to obtain with good accuracy but in this case, the standard
energy fluctuation method is quite effective. The use of the PDF 
complements traditional techniques utilizing energy fluctuations in 
Langevin simulations which are not suited to free energy and entropy 
calculations.

The QES nature of the DSHG theory allows not only the exact
computation of $E_0$ at several temperatures, but also of $\partial
E_0/\partial \beta$, using first order perturbation theory: $\partial
E_0/\partial \beta |_{\beta=\beta_0}=
(\Psi_0,\partial^2\Psi_0/\partial\phi^2)$ where 
$\beta_0$ is one of the special temperatures {\em e.g.}, 
Eq. (\ref{exactgs}). Thus the internal energy
$U$ and the entropy $S$ can also be found exactly at these
temperatures \cite{hks}. Once again, these quantities can be used to 
validate numerical work over a broad range of temperatures.

As a final point, we consider the relationship of the DSHG theory to
the more familiar Landau-Ginzburg model. Scrutiny of Eqs.
(\ref{vdshg})-(\ref{cosh}) reveals the following important
connection between the kink (and kink lattice) solutions of the
$\phi^4$ model and the double sine-Gordon (DSG) and DSHG
models. Consider the $\phi^4$ potential $V_4(u) =
[(n+\zeta)u^2-(n-\zeta)]^2$.  The substitution $u=\tanh\phi$ takes the
(static) equations of motion over to the DSHG equations. The alternative
substitution $u=\tan\phi$ leads to the DSG model. This means that {\em
all} known solutions of the $\phi^4$ theory can be directly taken over
to the DSHG and DSG theories (and vice versa). As one use of this
interesting relationship, the DSG kink lattice solution (not known
heretofore in the literature) can be written down directly for 
$V_{DSG}=(\zeta\cos 2\phi-n)^2$:
\begin{equation}
\phi_L=\pm\tan^{-1}\left(\tan\phi_1~\hbox{sn}
\left({x-x_0\over\xi_L},k\right)\right)~,
\end{equation}
simply by using the substitution $\tanh\rightarrow\tan$ in
Eq. (\ref{latt}). 

This connection enables us to write down by inspection not just the
kink solutions but their total energy as well, which is often a very
tedious task. Moreover, since we know that the DSHG model is an
example of a QES system, and considering the very similar way in which
the DSG and DSHG models are related to $\phi^4$, it is logical to
conjecture that the DSG model must also be a QES system. Indeed, this
is the case, and we have found several exact eigenvalues and
eigenfunctions for many temperatures.  The exact statistical
mechanical results for the DSG model, similar to the DSHG results
presented here, will be reported later \cite{hks}.

AK thanks Los Alamos National Laboratory for hospitality. SH
acknowledges useful discussions with Grant Lythe. The large-scale
simulations were performed on the CM-5 and Origin 2000 at the ACL, LANL, 
and on the T3E at NERSC, LBNL.

\end{document}

%% file: title2.tex
\catcode`\@=11

\def\maketitle2{\par 
\begingroup
\let\cite\@bylinecite
\def\thefootnote{\fnsymbol{footnote}}%
\twocolumn[\@maketitle2\vskip2pc]%
\thispagestyle{plain}\@thanks
\endgroup
\def\thefootnote{\arabic{footnote}}%
\setcounter{footnote}{0}%
\let\maketitle2\relax \let\@maketitle2\relax
\let\@thanks\relax \let\@authoraddress\relax \let\@title\relax
\let\@date\relax \let\thanks\relax \let\@abstract\relax 
\let\@pacs\relax}

\def\abstract#1{\gdef\@abstract{{\par 
\bgroup
\ifdim\prevdepth=-1000pt \prevdepth0pt\fi
\hsize\columnwidth
\dimen0=-\prevdepth \advance\dimen0 by17.5pt \nointerlineskip
\small\vrule width 0pt height\dimen0 \relax}{~~}#1\egroup}}

\def\pacs#1{\gdef\@pacs{{\par 
\bgroup
\hsize\columnwidth \parindent0pt
\ifdim\prevdepth=-1000pt \prevdepth0pt\fi
\dimen0=-\prevdepth \advance\dimen0 by20pt\nointerlineskip
\egroup} PACS numbers:~#1}}

\def\@maketitle2{
\@preprint
\@title
\ifdim\prevdepth=-1000pt \prevdepth0pt\fi
\@authoraddress
\@date
\begin{list}{}{\leftmargin=0.10753\textwidth \rightmargin=\leftmargin
\itemsep=1pc\partopsep=-1pc}
\item\@abstract
\item\@pacs
\end{list}
}

\catcode`\@=12